\documentclass[prl,preprintnumbers,twocolumn,floatfix,nofootinbib,letterpaper,superscriptaddress]{revtex4-2}
\usepackage[acronym]{glossaries}

\usepackage{graphicx}
\usepackage{bm}
\usepackage{amsmath,amssymb}
\usepackage{booktabs}
\usepackage{color}
\usepackage{units}

\makeglossaries
\newacronym{tbs}{TBS}{templated background search}
\newacronym{cbc}{CBC}{compact binary coalescence}
\newacronym{snr}{SNR}{signal-to-noise ratio}
\newacronym{psd}{PSD}{power spectral density}

\usepackage[normalem]{ulem}

\newcommand{\nrsur}{\texttt{NRSur7dq4}}
\newcommand{\pvt}{\texttt{IMRPhenomPv2}}
\newcommand{\xphm}{\texttt{IMRPhenomXPHM}}
\newcommand{\nrh}{\texttt{NRHybSur3dq8}}
\newcommand{\xas}{\texttt{IMRPhenomXAS}}
\newcommand{\seob}{\texttt{SEOBNRv5PHM}}

\usepackage{orcidlink}

\hypersetup{colorlinks=true, citecolor=teal, urlcolor=teal, linkcolor=teal}

\usepackage{lineno}
\begin{document}

\title{Distinguishing cosmology, astrophysics and systematics: the Bayesian search for a primordial gravitational-wave background in next-generation observatories}

\author{Teagan A.~Clarke~\orcidlink{0000-0002-6714-5429}} \email{teagan.clarke@princeton.edu}
\affiliation{Department of Physics, Princeton University, Princeton, New Jersey, 08544, USA}

\author{Sylvia Biscoveanu~\orcidlink{0000-0001-7616-7366}} \email{sbisco@princeton.edu}
\affiliation{Department of Physics, Princeton University, Princeton, New Jersey, 08544, USA}

\author{Carl-Johan Haster~\orcidlink{0000-0001-8040-9807}}
\email{carl.haster@unlv.edu}
\affiliation{Department of Physics and Astronomy, University of Nevada, Las Vegas, Nevada, 89154, USA}
\affiliation{Nevada Center for Astrophysics, University of Nevada, Las Vegas, Nevada, 89154, USA}

\date{\today}

\begin{abstract}
Next-generation gravitational-wave observatories like Cosmic Explorer will provide an unprecedented opportunity to probe the cosmological stochastic gravitational-wave background, providing insight into the early Universe that is inaccessible with other means. 
We investigate the prospects for measuring a Gaussian cosmological gravitational-wave background in Cosmic Explorer in the presence of a distinguishable, non-Gaussian foreground of binary black hole mergers, using a Bayesian search that simultaneously fits binary black hole and background parameters. 
We successfully recover a Gaussian cosmological background with a \acrlong{snr} of 5 in one day of simulated data in Cosmic Explorer without contamination from unresolved signals or data exclusion. 
However, when introducing waveform systematic errors to our analysis, our inference on the stochastic background parameters becomes biased and can falsely infer the presence of a background when there is none. 
The bias is driven by signal mismatches at low frequencies ($<\unit[20]{Hz}$), especially when the recovery waveform is missing physical effects included in the injected signals. 
While we propose a metric for diagnosing if measurements will be strongly biased by waveform systematics, our results suggest improvements in waveform modeling or other mitigation techniques will be required to accurately recover a primordial stochastic gravitational-wave background in next-generation detectors.
\end{abstract}

\maketitle

\textbf{\textit{Introduction.---}}
Detecting a primordial gravitational-wave background would allow us to observe and characterize the Universe at very early times that are out of reach of electromagnetic observatories.  
Although the background predicted from standard inflation models is too weak for all but the most ambitious proposed space-based observatories~\citep{Turner1997, Crowder2005, Kawamura2006, Corbin2006}, primordial backgrounds produced by alternate models like strongly first-order phase transitions \citep{Kosowsky_1992}, cosmic strings~\citep{Maggiore2000} or preheating~\citep{Khlebnikov1997, Easther2007} may be observable in next-generation ground-based observatories like Cosmic Explorer or Einstein Telescope~\citep{CE, ET}.

A Gaussian stochastic background signal will be obscured by a non-Gaussian astrophysical foreground of resolved and unresolved compact binary mergers. 
A careful strategy is needed to accurately disentangle a stochastic background signal from this astrophysical foreground. 
One option is to remove the foreground signals before searching the remaining data for the stochastic background by subtracting a best-fit waveform template~\citep[e.g.,][]{Cutler2006, Cutler2009,  Wu2012, Regimbau2017, Sachdev2020, Zhou2023, Pan2023, Kuwahara2026} or ``notching'' in time-frequency space~\citep{Zhong2024, Zhong2025, Zhong2026}.
However, difficulties with these methods arise if there are unresolved mergers or overlapping signals within the data, which must be accounted for with additional modeling techniques~\citep[e.g.,][]{Kuwahara2026}. 
Other efforts to simultaneously measure foregrounds and backgrounds employ a filtered component separation procedure, allowing for inference of different backgrounds with different spectral shapes simultaneously, but not accounting for the non-Gaussianity of the astrophysical foreground~\citep[e.g.,][]{Ungarelli2004, Parida2016, Poletti2021, Wang2025}.
\paragraph{}
The \acrfull{tbs}~\citep{Thrane2013, Smith2018, HernandezVivanco2019, Smith2020, Banagiri2020, Kou2025, Bers2026} is a Bayesian and statistically optimal search method for a non-Gaussian astrophysical background that simultaneously characterizes resolved and unresolved binary mergers.
The \acrshort{tbs} method has been extended to simultaneously resolve astrophysical and primordial backgrounds in the context of current ground-based observatories~\citep{Biscoveanu2020}, where the search has been demonstrated to avoid residual contamination without requiring any data to be excluded. 
\paragraph{}
Next-generation ground-based observatories like Cosmic Explorer or Einstein Telescope will be promising facilities for detecting a possible primordial gravitational-wave background with amplitude $\Omega_\alpha < 10^{-10}$. 
However, waveform model accuracy is known to require substantial improvement to achieve unbiased individual-event parameter estimation at the sensitivities projected for these future observatories~\citep[e.g.,][]{Purrer:2019jcp, Kapil2024}, which could introduce contamination when measuring a stochastic gravitational-wave background.  
For the subtraction and notching methods, residual power from imperfectly subtracting a point estimate for the foreground signal and waveform modeling uncertainties are limiting factors to the sensitivity to a primordial gravitational-wave background~\citep{Zhou2022, Zhou2023, Song2024}. 
Efforts towards mitigating the excess un-subtracted power include using e.g., a Fisher matrix approach to project away residuals~\citep{Harms2008, Sharma2020, Yamanoto_2026}. 
\paragraph{}
In this \textit{Letter}, we extend the \acrshort{tbs} method to search for cosmological backgrounds in next-generation ground-based observatories like Cosmic Explorer. At its proposed design sensitivity, Cosmic Explorer will resolve every binary black hole merger of stellar origin in the observable Universe and almost every neutron star merger~\citep{Regimbau2017,  Sachdev2020, Evans2021}, resulting in a detected binary black hole merger approximately every 200 seconds~\citep{Abbott2018}. 
We demonstrate, using a simulated population of binary black hole mergers in approximately one day of data, that the parameters of a primordial background can be inferred simultaneously with the fraction of segments that contain a binary black hole merger (referred to as ``duty cycle'' throughout) and binary black hole parameters. 
However, we also show that waveform systematics can strongly bias measurements of the background, with unmodeled signal power leaking into the inference of the background parameters. 
Finally, we explore some mitigation strategies to obtain unbiased inference of a background and discuss possible techniques to account for waveform errors during the \acrshort{tbs} procedure. 

\vspace{10pt}
\textbf{\textit{The \acrlong{tbs}.---}}
We seek to simultaneously fit the parameters of binary black hole mergers and the primordial gravitational-wave background using the \acrfull{tbs} method~\citep{Smith2018, Biscoveanu2020}. 
We divide the gravitational-wave data into segments of $\unit[T]{s}$ in duration. 
Each segment contains a Gaussian gravitational-wave background signal with either Gaussian noise, or Gaussian noise plus a binary black hole merger. 
The Gaussian background signal has a dimensionless energy density characterized by
\begin{equation}
    \Omega_\text{gw}(f) = \Omega_\alpha \left(\frac{f}{f_\text{ref}}\right)^\alpha,
\end{equation}
where $\alpha$ is the power law index, $\Omega_\alpha$ is the amplitude and $f$ is the gravitational-wave frequency. 
For each segment $i$, the likelihood of observing frequency-domain strain ($s_{i,k}$) in the frequency band $k$, comprised of a signal characterized as $h(\theta)$: a binary black hole merger with parameters $\theta$ and primordial background parameters $\{ \Omega_\alpha, \alpha \}$ is given by~\citep[e.g.,][]{Romano2017}
\begin{widetext}
\begin{equation}
    \mathcal{L}(s_{i,k}|\theta, \Omega_\alpha, \alpha ) = 
    \frac{1}{\text{det}(\pi T \textbf{C}_k(\Omega_{\alpha}, \alpha) / 2 )} \text{exp} \bigg( -\frac{2}{T} (s_{i,k}-h_k(\theta))^*\text{\textbf{C}}_k(\Omega_\alpha, \alpha)^{-1}(s_{i,k}-h_k(\theta)) \bigg),
    \label{eq:likelihood}
\end{equation}
where $\textbf{C}_k$ is the frequency-dependent covariance matrix for the segment, which for a two-detector network is given by
\begin{align}
    \textbf{C}_k = 
    \begin{pmatrix}
  \text{PSD}_1(f_k) + \kappa_{11}(f_k)\Omega_\text{gw} & \kappa_{12}(f_k)\Omega_\text{gw}\\ 
   \kappa_{21}(f_k)\Omega_\text{gw} & \text{PSD}_2(f_k) + \kappa_{22}(f_k)\Omega_\text{gw}
\end{pmatrix},
\label{eq:c-matrix}
\end{align}
\end{widetext}
with contributions from the detector \acrlong{psd} ($\text{PSD}_I(f_k)$). The quantity $\kappa_{IJ}$ converts $\Omega_\text{gw}$ into a strain \acrlong{psd} as a function of the detector overlap reduction function $\gamma_{IJ}$ and Hubble constant $H_0$:
\begin{align}
    \kappa_{IJ}(f_k) = \gamma_{IJ}(f_k)\frac{3H_0^2}{10\pi^2f^3}.
\end{align}
By defining separate marginalized likelihoods, the signal likelihood $\mathcal{L}_S$ and the noise likelihood $\mathcal{L}_N$ ($h_k(\theta)=0$), we can write the combined likelihood for $N$ segments, marginalized over the binary parameters and given a duty cycle $\xi$ as 
\begin{align}
    \mathcal{L}(\textbf{s}|\Omega_\alpha,\alpha,\xi) = \prod^N_i \xi\mathcal{L}_S(s_i|\Omega_\alpha,\alpha) + (1-\xi)\mathcal{L}_N(s_i|\Omega_\alpha,\alpha).
    \label{eq:likelihood_duty_cycle}
\end{align}
Additional details on our statistical framework are provided in the Supplemental Material.
We use this framework to measure the stochastic background parameters in a fiducial binary black hole population with different levels of contamination from waveform systematics. 

\vspace{10pt}
\textbf{\textit{Recovering a background with no systematics.---}}
We begin by demonstrating the \acrshort{tbs} framework in approximately one day of simulated Cosmic Explorer data. 
Assuming binary black holes merge every 640 seconds~\citep[e.g.,][]{TheLIGOScientificCollaboration2026}, we choose a merger duty cycle of 0.05 using $\unit[32]{s}$ segments. 
This amounts to 150 segments out of 3000 containing a merger. 
We tune the masses of the binary black holes such that the signals fit within $\unit[32]{s}$ of data from $\unit[5]{Hz}$, and draw from a power law in redshifted chirp mass with spectral index -3 ranging between $\unit[35]{M_\odot}$ and $\unit[200]{M_\odot}$. 
We draw from a Madau-Dickinson distribution~\citep{Madau2014} in redshift to a maximum of $z=5$.
Additional details on our fiducial population are provided in the Supplemental Material. 

Each signal is injected into Gaussian noise colored by the power spectral density for the proposed Cosmic Explorer observatory~\citep{CE}. 
 We assume two Cosmic Explorer instruments, one $\unit[40]{km}$ and one $\unit[20]{km}$, located at the sites of LIGO Hanford and Livingston.\footnote{We use the baseline power spectral density curves for Cosmic Explorer $\unit[40]{km}$ and $\unit[20]{km}$ instruments available at https://dcc.cosmicexplorer.org/CE-T2000017-v7, shown as the curves labeled ``40:CB'' and ``20:CB'' respectively in Figure~2 of Ref.~\cite{Srivastava2022}. }
All of the mergers are resolved in our Cosmic Explorer network, ranging in network matched-filter \acrfull{snr} from 8---1050.
Figure~\ref{fig:SNR} in the Supplemental Material shows the distribution of \acrshortpl{snr} we obtain.  
We infer the parameters of each signal using the \texttt{Bilby} inference library~\citep{bilby, Romero-Shaw:2020:Bilby} with the \texttt{dynesty} nested sampler~\cite{Speagle2020}. 
Then, we calculate the marginalized signal and noise likelihoods (Equations~\ref{eq:marg_S} and~\ref{eq:marg_N}) in post-processing. 
Details of this post-processing step are described in the Supplemental Material. 
Initially, we assume there are no waveform systematic errors by injecting and recovering with the \texttt{IMRPhenomXPHM} waveform model~\cite{Pratten2021_xphm, Colleoni2025_xphm}. 
We assume the population properties of merging binary black holes are known, leaving the development of a method for simultaneous inference of the population hyper-parameters and stochastic background parameters to future work. 

To begin, we analyze 3000 segments with the configuration described above \textit{in the presence of a Gaussian background.}
The amplitude and spectral index of the background are tuned such that the \acrshort{snr} of the background is 5: $\Omega_\alpha = 10^{-10.07}, \alpha=0$.
We note that while on the more optimistic side of current estimates for a primordial background~\citep{Giblin2014}, a background with these parameters has not been ruled out by searches in LIGO-Virgo-KAGRA data~\citep{Abbott2021, TheLIGOScientificCollaboration2025}. 
For the segments containing a binary black hole signal, we generate the waveforms from $\unit[5]{Hz}$ to ensure all higher mode content up to $\ell=4$ is in-band by the beginning of the likelihood integration at $\unit[10]{Hz}$. 

\begin{figure*}[!tbp]
    \centering
    \includegraphics[width=0.45\linewidth]{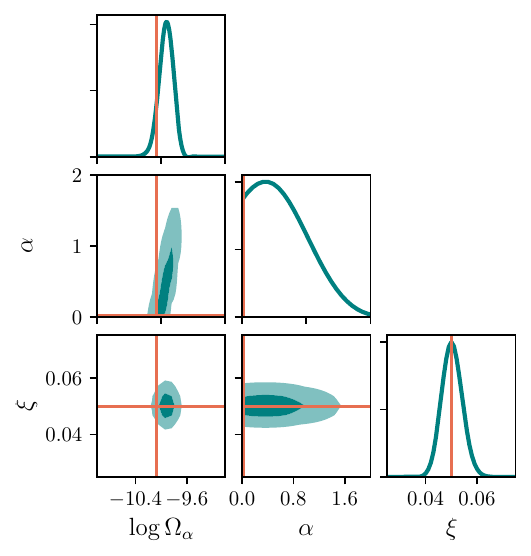}
    \includegraphics[width=0.45\linewidth]{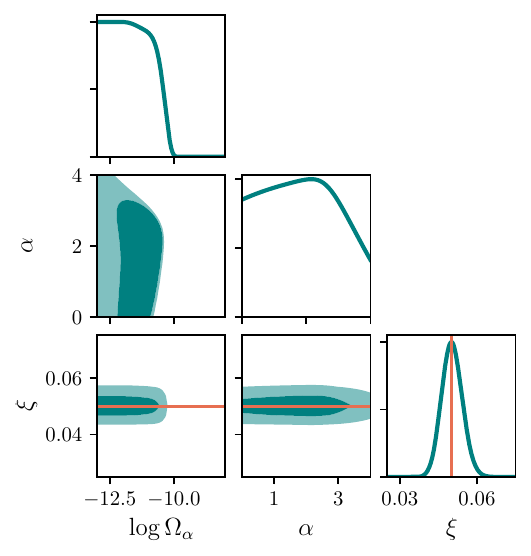}
    \caption{Marginalized posteriors for the recovered parameters of a simulated stochastic gravitational-wave background recorded after approximately one day of data in Cosmic Explorer. The shaded regions enclose the 50\% and 90\% credible regions of the 2D marginal posteriors. We accurately recover the injected amplitude and spectral index of the background as well as the binary merger duty cycle (shown as the orange solid lines). The left-hand side shows our results when we include a background signal with a combined \acrshort{snr} of 5 in 3000 segments, while the right-hand panel shows our inference for the lack of a background with 3000 segments.  }
    \label{fig:Stoch_XPHM}
\end{figure*}

The left-hand panel of Figure~\ref{fig:Stoch_XPHM} shows our results for the combined dataset. 
We successfully recover the injected amplitude, spectral index and duty cycle and measure a natural log Bayes factor in support of a background of 9.9, which is consistent with the expected scaling of $\ln \mathcal{B} \approx \text{SNR}^2/2$. 
We also test our framework's ability to infer the absence of a stochastic background over the same data duration and duty cycle. 
In this case, shown in the right-hand panel of Figure~\ref{fig:Stoch_XPHM}, we successfully rule out the presence of a background, obtaining $\ln \mathcal{B} = -1.5$ in support of the background hypothesis, and place an upper limit on the amplitude of the background of $\text{log}_{10} \Omega_\alpha < -10.9$ at the 90\% credibility level. 

\vspace{10pt}
\textbf{\textit{Impact of waveform systematics.---}}
Next, we investigate the impact of waveform systematic errors on our inference of the background parameters.
The worst-case scenario for accurately inferring a background would occur when there are very significant waveform systematics and a very high duty cycle. 
We simulate this case by injecting and recovering our binary black hole population with a range of mismatched waveform models including different physical effects, like spin precession and radiation in higher-order multipoles: \nrsur~\citep{Varma2019_nrsur}, \nrh~\citep{Varma2019_nrhyb}, \seob~\citep{Khalil2023_seob, Pompili2023_seob, vandeMeent2023_seob, Ramos-Buades2023_seob}, \xphm~\citep{Pratten2021_xphm, Colleoni2025_xphm}, \xas~\citep{Pratten2020_xas} and \pvt~\citep{Hannam2014_pv2, Khan2016_pv2, Husa2016_pv2}. 
We describe these waveform models in more detail in the Supplemental Material.

By mixing and matching different injection and recovery waveforms, we simulate different levels of waveform systematic error. 
We simulate a duty cycle of $\xi=1$ (i.e., all segments contain a binary black hole signal) to efficiently measure the effect of waveform systematics without requiring many times the number of segments analyzed. 
We do not inject a stochastic background for these investigations, and test how well we recover the absence of a background. 
Recovering spuriously strong support for the presence of a background would indicate that contamination from waveform systematics is a significant source of error in the accurate measurement of a background. 

\begin{figure*}
    \centering
    \includegraphics[width=\linewidth]{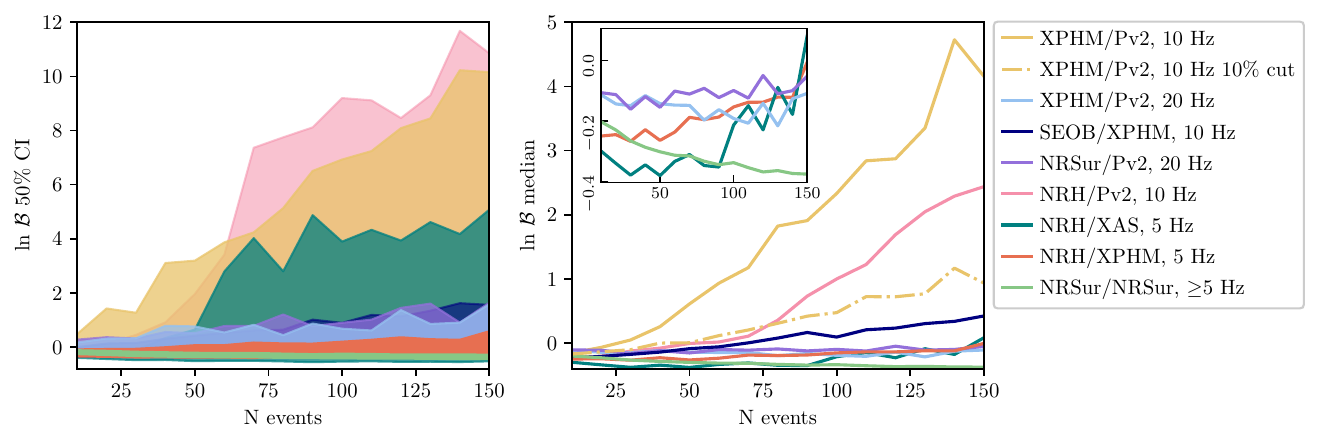}
    \caption{Natural log Bayes factor in support of the presence of a stochastic background for a variety of waveform and data conditioning combinations as a function of number of events included in the calculation. Left panel: The 50\% credible interval on the natural log Bayes factor for each of our tests after drawing 500 realizations of our population of 200 events. Right panel: the median natural log Bayes factor of each of out tests. Only analyses that start at $\unit[20]{Hz}$ (light blue and purple curves), employ waveform pairs with similar physics (orange, teal curves) or are absent of any systematics (lime green curve) avoid a strongly biased measurement in support of a background after 200 segments. For the most highly contaminated test (injecting with \texttt{IMRPhenomXPHM}, recovering with \texttt{IMRPhenomPv2}), we also show the median natural log Bayes factor when we exclude the worst ten percent of events according to the distinguishability statistic. }
    \label{fig:boostrap_median}
\end{figure*}

Table~\ref{tab:wf_bfs} lists each of the waveform pairs we consider and the natural log Bayes factor in support of a background obtained when combining 200 segments. 
Figure~\ref{fig:boostrap_median} shows the natural log Bayes factor in support of a stochastic background as more signal segments are added to our analysis for a variety of waveform combinations.
To mitigate against stochastic variance in the events selected at each stage, we sample with replacement over 500 realizations of event selection and then consider summary statistics of our sampling. 
In most of the tests, the natural log Bayes factor in favor of a background increases as more segments are included in the analysis. 
We show the 50\% credible interval on the left-hand side and the median on the right-hand side. 
For the tests that show the most contamination from waveform systematics, we find that the results are strongly impacted by individual outlier events. 
To understand this, we investigate the natural log Bayes factors obtained for a background for each single segment in our population for a selection of waveform pairs. 
This is shown in Figure~\ref{fig:single_Bfs} in the Supplemental Material. 
We find that in the case of waveform pairs that show strong contamination, there are one or two extreme outliers with $\ln \mathcal{B} >10\sigma$ above the mean of the distribution, that appear to drive most of the contamination, however we do not observe a clear correlation between these events' parameters and their individual Bayes factors.
As there would naturally be no individual segments that would produce such an extreme preference for a background, even when one is present, we interpret these events as signatures of un-physical contamination. 

\begin{table}
    \centering
    \begin{tabular}{l|l|l|c|c}
       Injection  & Recovery & Abbreviation & $f_\text{min}$ [Hz] & ln$\mathcal{B}$ \\
       \hline
        \texttt{IMRPhenomXPHM} & \texttt{IMRPhenomPv2} & XPHM/Pv2 & 10 & 6.5 \\
        \texttt{IMRPhenomXPHM} & \texttt{IMRPhenomPv2} & XPHM/Pv2 & 20 & -0.2 \\
        \texttt{SEOBNRv5PHM} & \texttt{IMRPhenomXPHM} & SEOB/XPHM & 10 & 0.5 \\
        \texttt{NRSur7dq4} & \texttt{IMRPhenomPv2} & NRSur/Pv2 & 20 & -0.1 \\
        \texttt{NRHybSur3dq8} & \texttt{IMRPhenomPv2} & NRH/Pv2 & 10 & 5.3 \\
        \texttt{NRHybSur3dq8} & \texttt{IMRPhenomXAS} & NRH/XAS & 5 & 0.7 \\
        \texttt{NRHybSur3dq8} & \texttt{IMRPhenomXPHM} & NRH/XPHM & 5 & -0.1 \\
        \hline
        \texttt{NRSur7dq4} & \texttt{NRSur7dq4} & NRSur/NRSur & $\geq5$ & -0.4 \\

    \end{tabular}
    \caption{Natural log Bayes factors in support of a background for 200 binary black hole signal segments analyzed with varying starting frequencies and waveform systematic severity. The most strongly biased measurements came from recovering with the $\ell=m=2$ mode-only waveform \pvt\ when injecting with a waveform that includes higher modes, when analyzing from $\unit[10]{Hz}$.  }
    \label{tab:wf_bfs}
\end{table}

\vspace{10pt}
\textbf{\textit{Strategies to avoid biased measurements.---}}
Our tests have shown that waveform systematics have the potential to strongly bias inference of a Gaussian background. 
As shown in Figure~\ref{fig:boostrap_median}, beginning the integration of the single-segment likelihood at $\unit[20]{Hz}$ is effective at guarding against the bias incurred by waveform systematics, indicating that most of the excess power is contaminating the background measurement at low frequencies, where our search is the most sensitive. 
However, only analyzing data $>\unit[20]{Hz}$ is sub-optimal, because obtaining an \acrshort{snr} of 5 with the background parameters simulated in Figure~\ref{fig:Stoch_XPHM} would require almost eight times the data length compared to integrating from $\unit[10]{Hz}$. 
We note that some waveform combinations return relatively unbiased results even when beginning the analysis at $\unit[5]{Hz}$ or $\unit[10]{Hz}$, most favorably injecting with \texttt{NRHybSur3dq8} and recovering with \texttt{IMRPhenomXPHM} (orange curves in Figure.~\ref{fig:boostrap_median}). 
We attribute this to the recovery waveform (\xphm) including all of the physical effects of \nrh, and the injected waveform omitting spin precession, the presence of which may be more likely to impart residual coherent signal in the data when recovering with a mismatched waveform model, since much of the difference between waveforms is encoded in the spin prescriptions. 

We seek to determine some criterion with which to predict if waveform systematics will seriously contaminate an analysis of a primordial gravitational-wave background. 
One rule of thumb to determine if two waveforms with mismatch $\mathcal{M}$ will be distinguishable at a certain \acrshort{snr} $\rho$ is given by~\citep[e.g.,][]{lindbolm_2008, baird_2013, Chatziioannou2017, Purrer:2019jcp}:
\begin{equation}
   \mathcal{M} \sim \frac{D_f}{2\mathrm{\rho^{2}}},
   \label{eq:SNR}
\end{equation}
where the mismatch between two waveforms $h_1$ and $h_2$ is given by
\begin{align}
    \mathcal{M} = 1- \frac{\langle h_1 | h_2 \rangle}{\sqrt{\langle h_1 | h_1 \rangle \langle h_2 | h_2 \rangle}},
\end{align}
and $D_f$ is the number of degrees of freedom that can impact the mismatch. 
The noise-weighted inner product $\langle h_1 | h_2 \rangle$ with \acrshort{psd} $S_n(f)$ is defined as \citep[e.g.,][]{Cutler_1994}
\begin{align}
    \langle h_1 | h_2 \rangle = 4\mathrm{Re} \int^{f_\text{high}}_{f_\text{low}} df \frac{h_1(f)h^*_2(f)}{S_n(f)}.
    \label{eq:inner_product}
\end{align}
In our case, $h_1$ and $h_2$ are the frequency-domain waveforms generated with the same parameters with the injection and recovery waveform model. 
We follow the procedure used in e.g., Ref.~\citep{Abac2025_231123} and minimize the mismatch over phase, time and polarization.
We adapt this rule-of-thumb guideline and propose the following \textit{distinguishability} statistic $\mathcal{D}$ as a proxy for predicting contamination from waveform systematics: 
\begin{equation}
    \mathcal{D} = \mathcal{M}\rho^2. 
    \label{eq:disting}
\end{equation}
This statistic $\mathcal{D}$ is approximately equal to $\langle h_1 - h_2 | h_1 - h_2 \rangle /2$, i.e., half of the squared \acrshort{snr} of the waveform residuals~\citep{McWilliams2010, Thompson2025}. 

\begin{figure}
    \centering
    \includegraphics[width=1.025\linewidth, trim={0.25cm 0.25cm 0.25cm 0.25cm},clip]{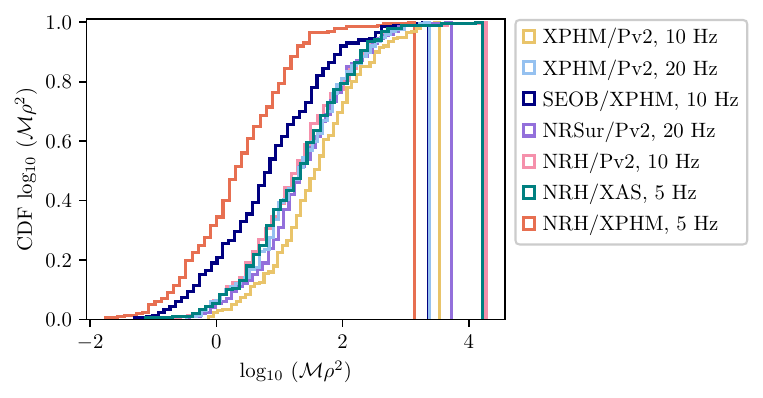}
    \caption{Cumulative density function for the distinguishability statistic for each of our waveform tests. The least to most biased measurements are displayed from left to right, demonstrating the usefulness of this statistic in identifying problematic measurements.}
    \label{fig:mm-cdf}
\end{figure}

We plot cumulative probability distributions of $\mathcal{D}$ for each of our waveform tests in Figure~\ref{fig:mm-cdf} and find that this statistic maps well to individual-event support for a background, with the most contaminated events corresponding to $\mathcal{D}>100$. 
The left and right-most curves (orange and yellow) in Figure~\ref{fig:mm-cdf} correspond to the least and most biased recoveries of the background, respectively. 
We also find improvement in the degree of bias incurred when omitting the tenth percentile highest segments in terms of the distinguishability statistic (including correcting for the imparted selection effects, described in further detail in the Supplemental Material), for the most strongly biased test in our sample: injecting with \xphm\ and recovering with \pvt. 
We show the new progression in natural log Bayes factor in the right-hand panel of Figure~\ref{fig:boostrap_median} in dot-dashed yellow. 
While the degree of bias is reduced, the support for a background still increases as more events are included in the analysis, indicating that a more systematic strategy for incorporating waveform systematic errors will be required for bias-free background inference at the sensitivity of next-generation detectors.

\vspace{10pt}
\textbf{\textit{Discussion and conclusions.---}}
In this \textit{Letter}, we demonstrate unbiased recovery of a primordial gravitational-wave background in the presence of an astrophysical foreground of binary black hole mergers at the sensitivity of a network of Cosmic Explorer observatories using one day of simulated data. 
Our results show that \acrshort{tbs} offers a safe, effective, and unbiased method for distinguishing Gaussian, cosmological backgrounds from the loud astrophysical foregrounds expected in next-generation observatories without residual contamination or excluding any data. 

However, we find that waveform systematics are important for accurate recovery of the background parameters, particularly for cases where frequencies below $20~\mathrm{Hz}$ are included in the analysis, and when using highly disparate waveform models that include different physical effects (e.g., \xphm\ and \pvt). 
While we observe a range of severity of the bias across waveform combinations, we note that for all waveform combinations, some realizations still return high support for a background at the 50\% credible level, such as the \xphm\ and \xas\ combination. 
Waveforms with any degree of mismatch show a bias that diverges from the matching waveforms case that could eventually result in a mis-modeled background, when including data with frequencies $\unit[<20]{Hz}$ in the analysis.  

Future analyses should account for waveform systematic errors during inference using, e.g., spline parameterizations~\citep[e.g.,][]{Edelman2021, Kumar2025} or more direct modeling, like marginalizing over the uncertainty in the fitting coefficients of semi-analytic waveform models~\citep[e.g.,][]{Mezzasoma:2025moh}. 
In this study, we have assumed that the population of binary black holes is known and sample from the same priors that we draw our population from. 
To avoid bias incurred by a mismatch between the true and assumed population in real data~\citep[e.g.,][]{Payne:2023kwj}, future studies should simultaneously infer the population parameters and the background parameters.
While we only analyze a high-mass population of mergers due to computational cost, we consider this to be a good proxy for the ``worst case scenario'', since lower-mass systems with longer inspirals should be in general less sensitive to waveform systematics due to having lower \acrshort{snr}, and correspondingly lower distinguishability (Equation~\ref{eq:disting}) on average~\citep[e.g.,][]{Hu2022}. 
Finally, our analysis should be expanded in the future to include contributions from binary neutron star and neutron-star black hole binaries, including possible overlapping signals due to their long inspiral time. 

\vspace{10pt}
\textbf{\textit{Acknowledgments.---}}
T.A.C and S.B are supported by NSF PHY-2513246.
C.-J.~H acknowledges the support from the Nevada Center for Astrophysics, from NASA Grant No. 80NSSC23M0104, and the NSF through Award No.~PHY-2409727.
Computing was performed on the Princeton University Della computing cluster. 

This work made use of the following software packages: \texttt{astropy} \citep{astropy:2013,astropy:2018,astropy:2022,astropy_21262391}, \texttt{Jupyter} \citep{2007CSE.....9c..21P,kluyver2016jupyter}, \texttt{matplotlib} \citep{Hunter:2007}, \texttt{numpy} \citep{numpy}, \texttt{pandas} \citep{mckinney-proc-scipy-2010,pandas_21500199}, \texttt{python} \citep{python}, \texttt{scipy} \citep{2020SciPy-NMeth,scipy_20764140}, \texttt{Bilby} \citep{bilby,Romero-Shaw:2020:Bilby, Bilby_21038138}, \texttt{dynesty} \citep{Speagle2020}, \texttt{corner.py} \citep{corner-Foreman-Mackey-2016,corner.py_21208839}, \texttt{Cython} \citep{cython:2011}, \texttt{h5py} \citep{collette_python_hdf5_2014,h5py_7568214}, \texttt{LALSuite} \citep{lalsuite,swiglal}, \texttt{sympy} \citep{sympy_15282092}, and \texttt{tqdm} \citep{tqdm_21624223}.
Software citation information aggregated using \texttt{\href{https://www.tomwagg.com/software-citation-station/}{The Software Citation Station}} \citep{software-citation-station-paper,software-citation-station-zenodo}.

\appendix
\section{Supplemental Material}

\textbf{\textit{Statistical details.---}}
In Equation~\ref{eq:likelihood}, we wrote down the likelihood for observing a gravitational-wave strain $s_{i,k}$ in a data segment $i$, in the frequency band $k$. 
For a segment that contains many frequency bins, the total likelihood is the product of the likelihood at each frequency bin;
\begin{equation}
    \mathcal{L}(s_i | \theta, \Omega_\alpha, \alpha ) = \prod_{k} \mathcal{L}(s_{i,k}|\theta, \Omega_\alpha, \alpha ).
\end{equation}

For each segment, we marginalize over the binary parameters $\theta$ and simultaneously infer the fraction of segments that contain a \acrlong{cbc} signal, $\xi$. To do this we use separate marginalized likelihoods, the signal likelihood $\mathcal{L}_S$ and the noise likelihood $\mathcal{L}_N$ which we write as
\begin{align}
    \mathcal{L}_S(s_i|\Omega_\alpha, \alpha) = \int d\theta \mathcal{L}(s_i|\theta,{\Omega_\alpha, \alpha})\pi(\theta),
\end{align}
and
\begin{align}
    \mathcal{L}_N(s_i | \Omega_\alpha, \alpha) = \mathcal{L}(s_i|\theta=0,{\Omega_\alpha, \alpha}).
    \label{eq:likelihood_N}
\end{align}
Finally, writing the combined likelihood from $N$ segments, marginalized over the binary parameters and given a duty cycle $\xi$, we obtain Equation~\ref{eq:likelihood_duty_cycle} from the main text. 

Calculating the marginalized likelihood written in Equation~\ref{eq:likelihood_duty_cycle} is computationally challenging since we need to compute the product over many individual likelihoods for each segment. 
We follow Ref.~\cite{Biscoveanu2020} and use likelihood reweighting~\citep{Payne2019} to evaluate the marginalized signal and noise likelihoods over a $[50\times50]$ grid in $\Omega_\alpha$ and $\alpha$ to obtain a three-dimensional marginalized likelihood. We marginalize over uniform priors in $\text{log}_{10}\Omega_\alpha$ and $\alpha$ between [-13,-8] and [0,4] respectively. 
Then, the marginalized signal likelihood for each segment becomes a Monte Carlo integral over the grid in $\Omega_\alpha$ and $\alpha$ and the $n$ posterior samples produced during inference: 
\begin{align}
    \mathcal{L}_S(s_i|\Omega_\alpha, \alpha) = \frac{\mathcal{Z}_{0,i}}{n}\sum^n_j \frac{\mathcal{L}(s_i|\theta_j, \Omega_\alpha, \alpha)}{\mathcal{L}(s_i|\theta_j, \Omega_\alpha=0)},
\end{align}
where $\mathcal{Z}_{0,i}$ is the evidence calculated during inference using the $\Omega_\alpha = 0$ likelihood, i.e.
\begin{align}
    \mathcal{Z}_{0,i} = \int d \theta \mathcal{L}(s_i|\theta, \Omega_\alpha=0)\pi(\theta).
\end{align}
Finally, we calculate the joint likelihood in Equation~\ref{eq:likelihood_duty_cycle} over a grid of 100 equally-spaced points in $\xi$ between 0 and 1. 

We can compute a Bayes factor to quantify the probability of the primordial Gaussian background hypothesis over the ``noise'', i.e., astrophysical foreground-only, hypothesis: 
\begin{align}
  \ln \mathcal{B} = \ln \mathcal{Z}_S - \ln \mathcal{Z}_N,
\end{align}
where $\mathcal{Z}_S$ and $\mathcal{Z}_N$ are the Bayesian evidence for a Gaussian background signal marginalized over the stochastic background parameters:

\begin{align}
    \mathcal{Z}_S = \int d\Omega_\alpha \ d\alpha \ d\xi  \mathcal{L}(\textbf{s}|\Omega_\alpha, \alpha, \xi)\pi(\Omega_\alpha, \alpha, \xi)
    \label{eq:marg_S}
\end{align}
and 
\begin{align}
    \mathcal{Z}_N = \int d\xi \prod_i^N \xi \mathcal{Z}_{0,i} + (1-\xi)\mathcal{Z}_{N,i},
    \label{eq:marg_N}
\end{align}
where $\mathcal{Z}_{0,i}$ is the noise evidence assuming no binary black hole signal and no stochastic background signal (Equation~\ref{eq:likelihood_N}, $\Omega_\alpha = 0)$, and $\mathcal{Z}_{N,i}$ is the likelihood from Equation~\ref{eq:likelihood} with $\Omega_\alpha=0,\ h(\theta)=0$. 

\begin{figure}
    \centering
    \includegraphics[width=\linewidth]{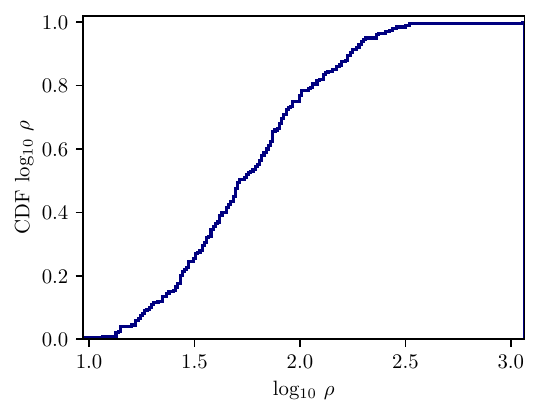}
    \caption{Cumulative density distribution of network matched-filter \acrlong{snr}'s obtained for our population of 200 binary black hole mergers in Cosmic Explorer. }
    \label{fig:SNR}
\end{figure}

\vspace{10pt}
\textbf{\textit{Details on simulated population of binary black hole mergers.---}}
We simulate a population of high-mass binary black hole mergers to ensure that all signals fit within $\unit[32]{s}$ of data from $\unit[5]{Hz}$. 
We draw from a power-law distribution in detector-frame chirp mass:
\begin{equation}
    p(\mathcal{M})\propto \mathcal{M}^{-3}
\end{equation}
between $\unit[35-200]{M_\odot}$, which is consistent with the current inferred astrophysical population of binary black hole mergers at high masses~\citep[e.g.,][]{TheLIGOScientificCollaboration2026}. We sample a uniform distribution in mass ratio ($q=m_2/m_1$, with $m_1 \geq m_2$) between 0.167 and 1. We draw from uniform distributions in the component spin magnitudes from 0 to 0.99 
and isotropic distributions in the spin orientations and sky locations. 
For the luminosity distance, we draw from the cosmic star formation history fit as a function of redshift $z$ from Ref.~\cite{Madau2014};
\begin{align}
    \psi(z) = \unit[0.015\frac{(1+z)^{2.7}}{1+[(1+z)/2.9]^{5.6}}]{M_\odot year^{-1}Mpc^{-1}},
\end{align}
with a maximum luminosity distance of $\unit[47]{Gpc}$, corresponding to a redshift of $\approx 5$ using Planck 2015 $\Lambda\text{CDM}$ cosmology~\citep{PlanckCollaboration2016}. 
We plot the resulting \acrshort{snr} distribution obtained when drawing 200 events from this population in Figure.~\ref{fig:SNR}

\vspace{10pt}
\textbf{\textit{Waveform details.---}}
Here we provide a brief description of the features and accuracy of each of the waveform models considered in our analysis. 
\begin{itemize}
    \item \texttt{IMRPhenomXPHM}~\citep{Pratten2021_xphm, Colleoni2025_xphm}: A phenomenological model that includes precession and radiation from higher-order multipoles ($(\ell, m)=$ (2,2), (2,1), (3,3), (3,2) and (4,4) modes) that produces good accuracy for generic precessing binaries across a broad parameter space but relies on an approximate precession-mapping using the ``twisting'' algorithm. 
    \item \texttt{SEOBNRv5PHM}~\citep{Khalil2023_seob, Pompili2023_seob, vandeMeent2023_seob, Ramos-Buades2023_seob}: An effective-one-body model that includes precession and higher multipoles ($(\ell, m)=$ (2,2), (2,1), (3,3), (4,4) and (5,5) modes) and is generally more accurate than \xphm\ due to an upgraded algorithm for precession dynamics and the addition of the (5,5) mode.  
    \item \texttt{NRSur7dq4}~\citep{Varma2019_nrsur}: A direct numerical-relativity surrogate for generic precessing binaries with mass ratios down to $q=m_2/m_1=0.2$, including higher multipoles ($(\ell, m)=$ (2,0), (2,1), (2,2), (3,0), (3,1), (3,2), (3,3), (4,2), (4,3) and (4,4) modes). It produces the highest accuracy within its calibrated parameter space but is more limited in the number of inspiral cycles it can generate than semi-analytic models like \xphm\ and \seob.
    \item \texttt{NRHybSur3dq8}~\citep{Varma2019_nrhyb}: A hybridized numerical-relativity surrogate for aligned-spin binaries with mass ratios down to $q=0.125$, including the same higher modes as \nrsur. It achieves very high accuracy for non-precessing systems and is able to model an arbitrary number of inspiral cycles.
    \item \texttt{IMRPhenomPv2}~\citep{Hannam2014_pv2, Husa2016_pv2, Khan2016_pv2}: A precessing phenomenological model using only the dominant $(\ell, m)=$(2,2) mode and a simple spin prescription using a single effective precession angle \citep{Schmidt_2015}, making it faster but less accurate than \xphm\ or \seob\,, particularly for highly precessing or asymmetric-mass systems.
    \item \texttt{IMRPhenomXAS}~\citep{Pratten2020_xas}: An aligned-spin phenomenological model using only the dominant $(\ell, m)=$(2,2) mode, with more refined calibration to numerical relativity than \pvt. It achieves high accuracy for equal-mass, non-precessing binaries but is inapplicable to systems with spin precession.
\end{itemize}

\begin{figure}
    \centering
    \includegraphics[width=\linewidth]{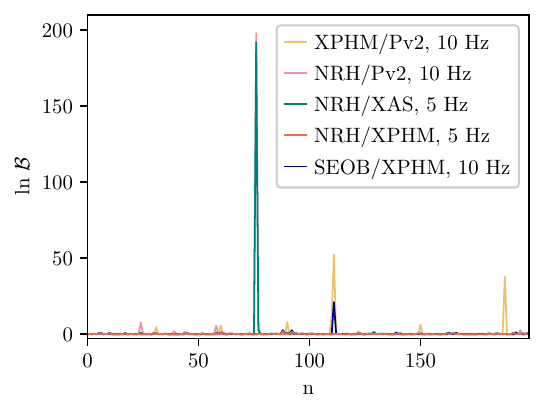}
    \caption{Individual natural log Bayes factors for the waveform combinations that show some level of contamination from waveform systematics. For most of the tests, there are a small number of extreme outliers that have a large impact on the overall analyses. }
    \label{fig:single_Bfs}
\end{figure}

\vspace{10pt}
\textbf{\textit{Investigation of outliers.---}}
Figure~\ref{fig:single_Bfs} shows the individual natural log Bayes factors obtained for each event in our population for five of the tests that showed any level of contamination from waveform systematics.
All of the pairs, except for \nrh/\xphm---which showed the least significant contamination---have extreme outliers that register support for a background that would be un-physical even if there were a background present. We interpret these outliers as red flags that can be used to help diagnose if a measurement is impacted by waveform systematics.

\vspace{10pt}
\textbf{\textit{Selection effects.---}} The cut to exclude the 10\% of segments with the highest distinguishability statistic in the analysis presented in the dashed-dot line in Figure~\ref{fig:boostrap_median} introduces a selection effect that must be accounted for in the likelihood. Following Ref.~\cite{Bers2026}, the individual-segment likelihood implied in Equation~\ref{eq:likelihood_duty_cycle} must be normalized when integrating over all data realization passing the selection criterion so that
\begin{align}
    &\mathcal{L}(s_i|\Omega_\alpha,\alpha,\xi, \mathrm{sel}) =\nonumber \\
  &\frac{1}{C(\xi, \Omega_\alpha,\alpha)} \left[\xi\mathcal{L}_S(s_i|\Omega_\alpha,\alpha) + (1-\xi)\mathcal{L}_N(s_i|\Omega_\alpha,\alpha)\right],\\
    &C(\xi, \Omega_\alpha,\alpha) = \int_{s_{\mathrm{sel}}}\left[\xi\mathcal{L}_S(s_i|\Omega_\alpha,\alpha) + (1-\xi)\mathcal{L}_N(s_i|\Omega_\alpha,\alpha)\right] ds.
    \label{eq:likelihood_selection_norm}
\end{align}
We can introduce two new variables,
\begin{align}
    \gamma(\Omega_\alpha,\alpha) = \int_{s_{\mathrm{det}}}\mathcal{L}_S(s_i|\Omega_\alpha,\alpha)ds = \frac{N_S^{\mathrm{sel}}}{N_S}, \label{eq:signal_norm}\\
    \beta(\Omega_\alpha,\alpha) = \int_{s_{\mathrm{det}}}\mathcal{L}_N(s_i|\Omega_\alpha,\alpha)ds = \frac{N_N^{\mathrm{sel}}}{N_N},
    \label{eq:noise_norm}
\end{align}
where $\gamma(\Omega_\alpha,\alpha)$ and $\beta(\Omega_\alpha,\alpha)$ represent the fraction of signal and noise segments that meet the selection criterion, respectively. 

In the last step of Eqs.~\ref{eq:signal_norm}-\ref{eq:noise_norm}, we have assumed that the probability of a given segment passing the selection criterion is independent of the stochastic background parameters. If the selection criterion depends primarily on the properties of the binary black hole merger in each segment, like the distinguishability criterion threshold we impose, this is a reasonable simplification, even though the strength of the Gaussian background changes the auto-power in each detector used as the \acrshort{psd} when calculating the inner product in Equation~\ref{eq:inner_product}.
By dropping the dependence on the stochastic background parameters $\{\Omega_{\alpha}, \alpha\}$, the normalization constant in Equation~\ref{eq:likelihood_selection_norm} can then be written as
\begin{align}
C(\xi) = \xi\gamma + (1-\xi)\beta,
\end{align}
where we use our simulated population to calculate the number of signal (noise) segments that meet the selection criterion, $N_{\mathrm{S(N)}}^{\mathrm{sel}}$.

\bibliographystyle{apsrev4-2}
\bibliography{bib}

\end{document}